\documentclass[preprint,aps,12pt,notitlepage,nofootinbib,tightenlines]{revtex4}
\usepackage{amsmath}
\usepackage{bm}
\usepackage{times}
\usepackage{braket}
\usepackage{color}
\usepackage{epsfig}
\usepackage{slashed}
\usepackage{hyperref}
\usepackage{multirow}
\usepackage{booktabs}
\usepackage{array}
\newcommand{\beq}{\begin{eqnarray}}
\newcommand{\eeq}{\end{eqnarray}}

\def \cpc{ {\bf Chin. Phys. C} }
\def \csb{ {\bf Chin. Sci. Bull.} }

\def \ijmpa{ {\bf Int. J. Mod. Phys. A}  }
\def \copc{ {\bf Comput. Phys. Commum. } }
\def \epjc{{\bf Eur. Phys. J. C} }
\def \jpg{ {\bf J. Phys. G} }
\def \npb{ {\bf Nucl. Phys. B} }
\def \plb{ {\bf Phys. Lett. B} }

\def \prd{ {\bf Phys. Rev. D} }
\def \prl{ {\bf Phys. Rev. Lett.}  }

\def \jhep{ {\bf JHEP}  }
\def \appb{ {\bf Acta Phys. Polon. B} }
\definecolor{Red}{rgb}{1.,0.,0.}

\definecolor{Blue}{rgb}{0.,0.,1.}

\definecolor{nicered}{rgb}{0.7,0.1,0.1}
\definecolor{nicegreen}{rgb}{0.1,0.5,0.1}
\def\lsim{ {\ \lower-1.2pt\vbox{\hbox{\rlap{$<$}\lower6pt\vbox{\hbox{$\sim$}}}}\ } }
\def\gsim{ {\ \lower-1.2pt\vbox{\hbox{\rlap{$>$}\lower6pt\vbox{\hbox{$\sim$}}}}\ } }
\hypersetup{colorlinks,citecolor=nicegreen,linkcolor=nicered}
\begin{document}
\title{Probing  anomalous top-Higgs couplings at the HL-LHC \\
via $H\to WW^{\ast}$ decay channels}
\author{Yao-Bei Liu\footnote{E-mail: liuyaobei@hist.edu.cn}}
\affiliation{Henan Institute of Science and Technology, Xinxiang 453003, P.R.China}
\author{Stefano Moretti\footnote{E-mail: s.moretti@soton.ac.uk}}
\affiliation{School of Physics \& Astronomy, University of Southampton, Highfield, Southampton SO17 1BJ, UK}
\vspace*{2truecm}
\begin{abstract}
We study the prospects of probing the anomalous $tHq$~($q= u, c$) couplings via SS2L or 3L signatures at the High Luminosity (HL-LHC) run of the 14 TeV CERN collider. We focus on signals of the  $tH$ associated production followed by the decay modes $t\to b\ell^{+}\nu_{\ell}$ and $H\to WW^{\ast}$, and $t\bar{t}$ production followed by the decay modes $t\to b\ell^{+}\nu_{\ell}$ and $\bar{t}\to H(\to WW^{\ast})\bar{q}$, where $\ell=e, \mu$. Based on two types of  $H\to WW^{\ast}$ decay topologies, one assuming the semileptonic decay mode $H\to WW^{\ast}\to \ell^{+}\nu jj $ and the other the fully leptonic decay mode $H\to WW^{\ast}\to \ell^{+}\nu \ell^{-}\bar{\nu}$, we perform a full simulation for
signals and backgrounds. It is shown that, at the future HL-LHC, the branching ratio $Br(t\to uh)~(Br(t\to ch))$ can be  probed to $1.17~(1.56)\times 10^{-3}$ for the same-sign di-lepton channel, and to $7.1~\times 10^{-4}~(1.39~\times 10^{-3})$ for the 3L channel at $3\sigma$ sensitivity.
\end{abstract}

\maketitle
\newpage
\section{Introduction}
 Processes mediated by Flavor Changing Neutral Currents (FCNCs) are very rare in
the Standard Model (SM) due to the Glashow-Iliopoulos-Maiani (GIM) mechanism~\cite{gim}.
However, because of the extended flavor structures existing in many New Physics (NP) models, the two-body FCNC decays $t\to qX$ ($q=u/c$ and $X=g/\gamma\/Z/H$) can be greatly enhanced: for example, in the Minimal Supersymmetric Standard Model~(MSSM) with branching ratio $Br(t\to cH)\sim 10^{-5}$ ~\cite{mssm}, in R-parity violating Supersymmetry (SUSY) with branching ratio
$Br(t\to cH)\sim 10^{-6}$~\cite{plb-2001-510}, in
2-Higgs-Doublet Models~(2HDMs) with branching ratio $Br(t\to cH)\sim 10^{-5}-10^{-3}$~\cite{2hdm}, in the little Higgs model with T-parity and the warped extra dimensions both with branching ratio $Br(t\to cH)\sim 10^{-5}$ ~\cite{lht,ED} and so on. Thus any experimental signatures of such FCNC processes
will serve as a clear signal for NP Beyond the SM (BSM) \cite{top-NP}. Up to now,  top-Higgs FCNC interactions have been studied widely via anomalous top decays or anomalous production processes of single top quark~\cite{t1,t2,t3,t4,prd86-094014,jhep-1407-046,150908149,prd89-054011,plb703-306}.

Currently, the ATLAS and CMS collaborations have carried out searches~\cite{cms-8,atlas-8,atlas13,cms13,1805.03483} for $tqH$ interactions with 7, 8
and 13 TeV data from the LHC.
For example, using 13 TeV data, the ATLAS and the CMS experiments have studied the $tqH$ FCNC processes in top quark pair events with $H\to \gamma\gamma$ for ATLAS and $H\to b\bar{b}$ for CMS. The resulting observed (expected) limits for $Br(t\to qH)$ at $95\%$ Confidence Level (CL) have been found to be~\cite{atlas13,cms13}:
\beq
Br(t\to Hu)\leq\left\{ \begin{array}{ll}
2.4~(1.7)\times 10^{-3} & {\rm ATLAS} \\
4.7~(4.3)\times 10^{-3} & {\rm CMS} \\ \end{array} \right. \nonumber \\
Br(t\to Hc)\leq\left\{ \begin{array}{ll}
2.2~(1.6)\times 10^{-3} & {\rm ATLAS} \\
4.7~(4.4)\times 10^{-3} & {\rm CMS} \\ \end{array} \right. \label{eq:tch}
\eeq
Very recently, a search for production of top pairs in which one top quark decays via
$t\to qH$ is reported by the ATLAS Collaboration~\cite{1805.03483}, with the subsequent Higgs boson decay to final states with at least one electron or muon.  The upper limits on the branching fractions $Br(t\to Hc)< 0.16\%$ and $Br(t\to Hu)< 0.19\%$ at $95\%$ CL are obtained~(with expected limits of $0.15\%$ in
both cases).
Apart from direct collider measurements, the upper limits of $Br(t\to qH) < 5 \times 10^{-3}$ and
$Br(t\to qH) < 0.21\%$ can be  obtained by bounding the $tqH$ vertex from the observed $D^{0}-\bar{D^{0}}$ mixing~\cite{prd81-077701} and $Z\to c\bar{c}$~\cite{prd72-057504}, respectively.

The upcoming project of the HL-LHC
is expected to reach 3 ab$^{-1}$. Preliminary sensitivity studies for the HL-LHC suggest the upper bound on $Br(t\to qH)$ to become about $1.5 \times 10^{-4}$ at $95\%$ CL by the ATLAS Collaboration~\cite{atlas-14-3000}. Further, many phenomenological studies within model-independent methods have been
performed from different channels~\cite{tfcnc-th,multilepton,prd-zhang,t5,t6,t7,t8,t9,t10}.
In this work, we study the prospects of probing the anomalous $tHq$ couplings by considering the processes of $tH$ associated production and $t\bar{t}$ production at the HL-LHC. We analyze two kinds of final states through leptonic top quark decays and $H\to WW^{\ast}$, one with Same Sign 2-Lepton~(SS2L) and the other with 3-Lepton (3L) topology, where the Higgs boson decays into a semi-leptonic $(H\to WW^{\ast}\to \ell^{+}\nu jj)$ or fully leptonic ($H\to WW^{\ast}\to \ell^{+}\nu\ell^{-}\bar{\nu}$) mode. The advantage of these channels is that their final states including the SS2L or 3L topologies can be used to significantly suppress QCD
backgrounds~\cite{ss2l}, which have not been fully studied in  previous literature.

The organization of this paper is as follows. In Sec.~II, we discuss two kinds of final states for the processes of $tH$ associated production with the decay chain $t\to W^{+}b\to \ell^{+}\nu b$ and $H\to WW^{\ast}$ as well as $t\bar{t}$ production with the decay chain $t\to \ell^{+}\nu_{\ell}b$ and $\bar{t}\to H(\to WW^{\ast})\bar{q}$. Then we discuss the HL-LHC sensitivity to the anomalous $tHq$ couplings. We summarize in Sec.~III.

\section{Numerical calculations and discussions}
 The general Lagrangian for  FCNC top interactions with the Higgs boson can be written as
\begin{equation}
{\cal L}= \kappa_{tuH}\bar{t}Hu+\kappa_{tcH}\bar{t}Hc+h.c.,
\label{tqh}
\end{equation}
{where the FCNC coupling parameters, $\kappa_{tuH}$ and $\kappa_{tcH}$, are real and symmetric since we do not consider here the CP violating effects}.

We perform systematic Monte Carlo (MC) simulations and study the sensitivity to the anomalous $tHq$ couplings through the associated $tH$ and $t\bar{t}\to tH\bar{q}$ processes at HL-LHC. We first extract the relevant  Feynman rules via the FeynRules package~\cite{feynrules} and generate the events with MadGraph5-aMC$@$NLO~\cite{mg5}. The signal and backgrounds samples are simulated at parton level with the NN23LO1 Parton Distribution Function~(PDF) set~\cite{cteq} and then passed through PYTHIA6.4~\cite{pythia} and DELPHES 3~\cite{delphes} for parton shower and detector
simulations, with the MLM matching scheme~\cite{MLM}  adopted.
Finally, event analysis is performed by using MadAnalysis5 \cite{ma5}.

\subsection{Analysis of the SS2L channel}
\begin{figure}[htb]\vspace{0.5cm}
\begin{center}
\centerline{\epsfxsize=16cm \epsffile{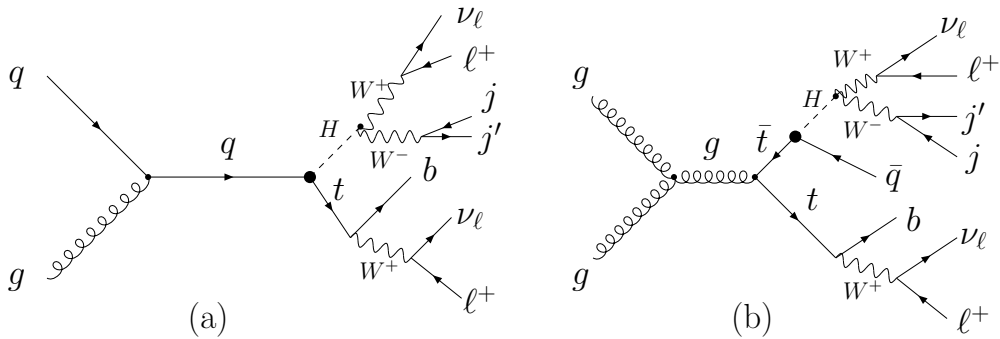}}
\vspace{-17cm}
\caption{Representative Feynman diagrams for the associated $tH$ process~(left) and the FCNC decay of the top pair production process~(right). Here $q=u,c$.}
\label{fey1}
\end{center}
\end{figure}

For the final states including the SS2L topology, the signals are generated through the following processes,
\beq\label{signal}
pp&\to& t(\to W^{+}b\to \ell^{+}\nu b)H(\to WW^{\ast}\to \ell^{+}\nu jj),\\
pp&\to& t(\to W^{+}b\to \ell^{+}\nu b)\bar{t}(\to Hq\to WW^{\ast}q\to \ell^{+}\nu jjq),
\eeq
where $\ell =e, \mu$. The representative Feynman diagrams are shown in Fig.~\ref{fey1}.

For this channel, the typical signal is exactly two same-sign leptons plus at least three jets, with at least one jet identified as $b$-jet, and missing transverse energy. The main backgrounds are $t\bar{t}V$ ($V=W, Z$), $W^{+}W^{+}jj$ and $W^{+}Zjj$. The $t\bar{t}$ process, which has large cross section, may also contribute to background if the a same-sign lepton pair comes from a $B$-hadron semi-leptonic  decay in the $b$-jet.
 We do not consider other backgrounds from $t\bar{t}H$, $t\bar{t}t\bar{t}$, tri-boson events and $tHj$. They are neglected because the cross sections are all negligible after applying  the selection cuts.

 The cross sections of dominant backgrounds at Leading Order (LO) are adjusted to Next-to-LO (NLO) by means of
$K$-factors, which are 1.04 for $W^{+}W^{+}jj$ jets~\cite{wwjj}, 1.24 for $t\bar{t}W$~\cite{nlo-ttw} and 1.39 for $t\bar{t}Z$~\cite{nlo-ttz}. The dominant  $t\bar{t}$ background is normalized to the NNLO QCD cross section of
953.6 pb~\cite{1303.6254}. For the $tH$ production cross section, the K-factor is taken as 1.5 at the 14 TeV LHC~\cite{prd86-094014}.

The decay chain $H\to WW^{\ast}\to \ell \nu jj$ may result in soft leptons and light jets, especially when they are coming from an off-shell $W$ boson. To analyze the signal sensitivity, we thus employ the
following basic cuts to select the events:
 \begin{itemize}
\item
Basic cuts: $p_{T}(\ell) > 10 \rm ~GeV$, $p_{T}(j, b) > 15 \rm ~GeV$, $|\eta_{\ell, j, b}|<2.5$,  where $\ell=e, \mu$.
\end{itemize}

\begin{figure}[htb]
\begin{center}
\centerline{\hspace{2.0cm}\epsfxsize=11cm\epsffile{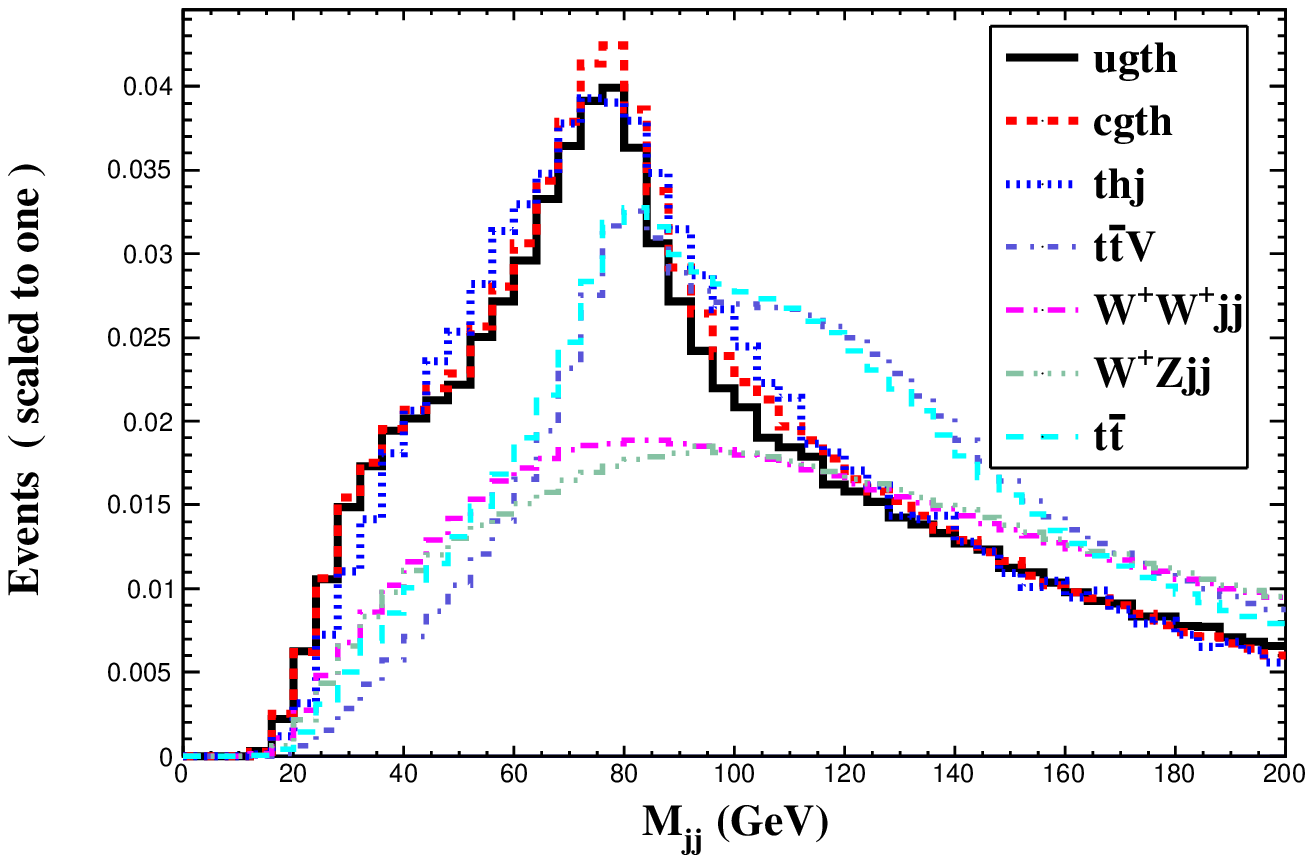}
\hspace{-3.0cm}\epsfxsize=11cm\epsffile{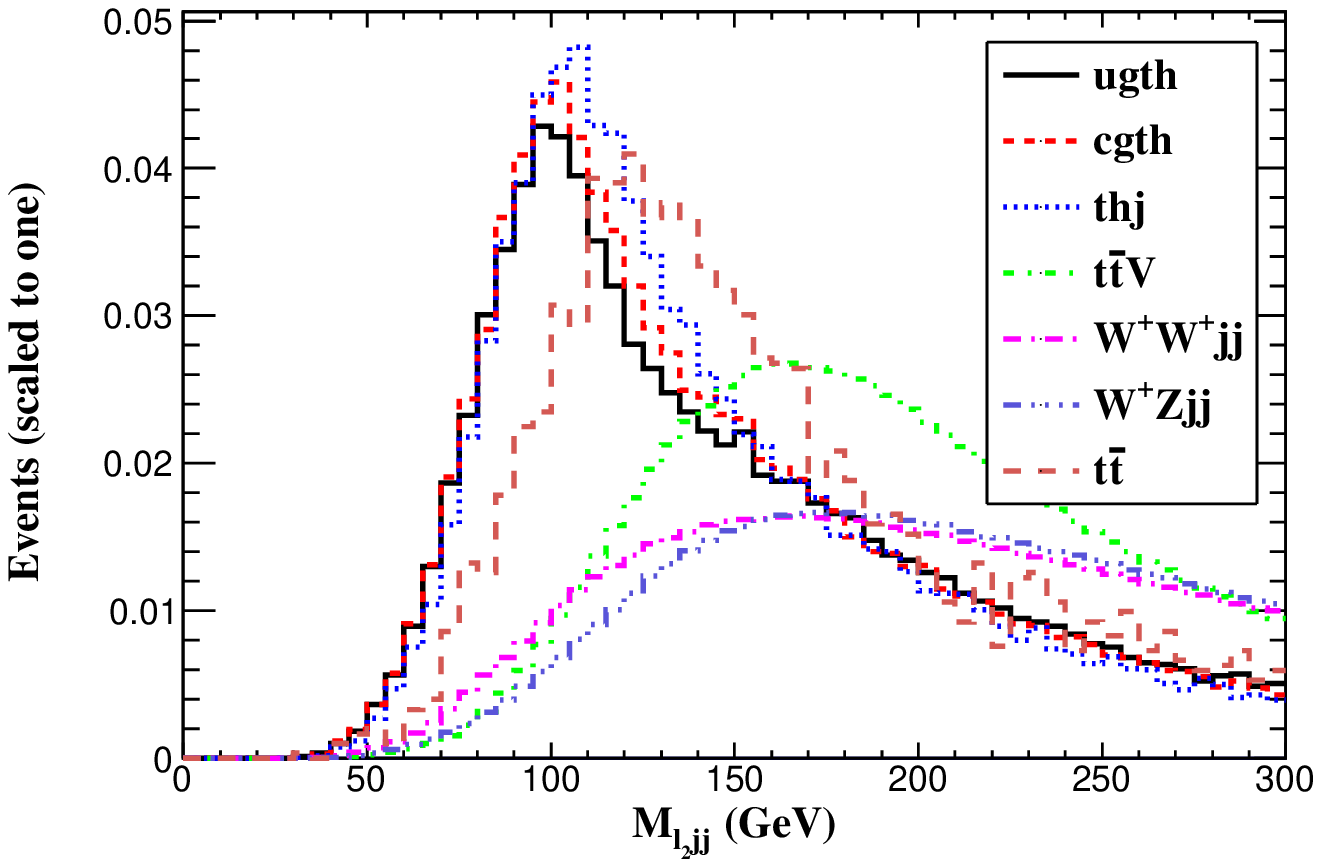}}
\centerline{\hspace{2.0cm}\epsfxsize=11cm\epsffile{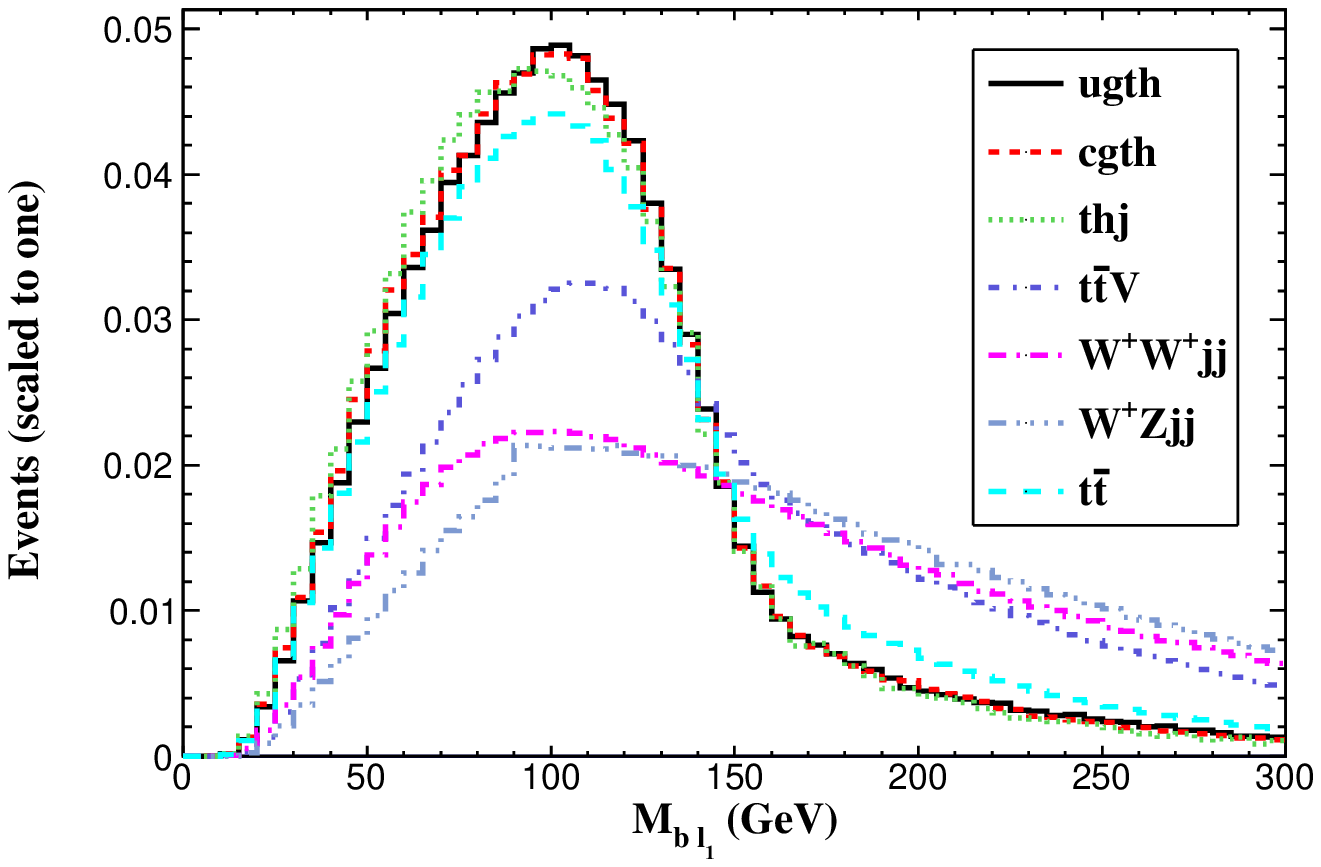}
\hspace{-3.0cm}\epsfxsize=11cm\epsffile{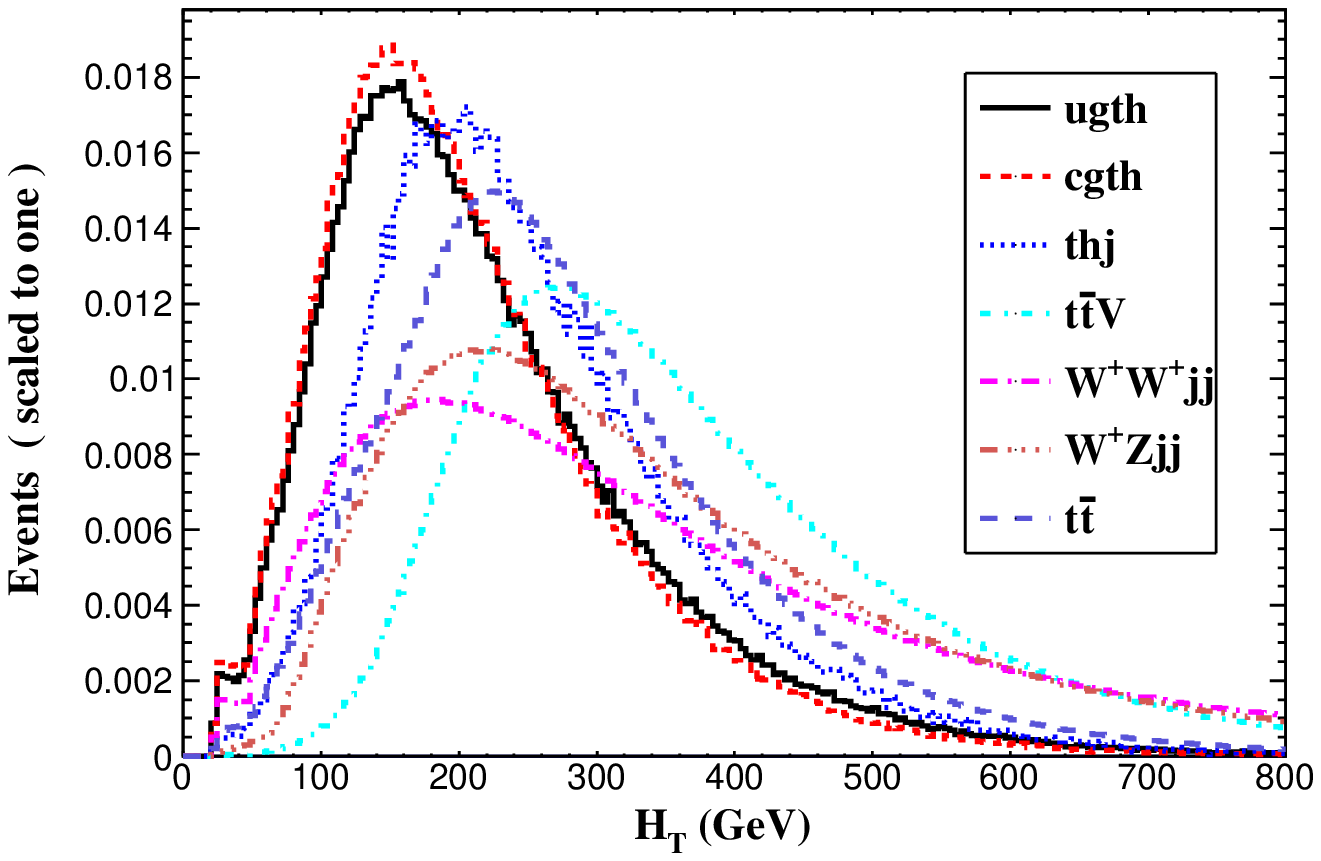}}
\caption{Normalized distributions for the signals and the backgrounds.}
\label{th-2l}
\end{center}
\end{figure}

In order to choose appropriate kinematic cuts, we plot in Fig.~\ref{th-2l} examples of kinematic distributions for the signal and backgrounds. Based on these distributions, we impose a further set of cuts.
 \begin{itemize}
\item
Cut-1: Exactly two same-sign leptons ($N(\ell^{+})=2$) with $p_T(\ell_{1})>20$ GeV and $p_T(\ell_{2})>10$ GeV~($\ell_{1}$ and $\ell_{2}$ denote the higher and lower $p_T$ lepton, respectively) plus exactly one $b$-tagged jet ($N(b)= 1$). To remove contamination from hadron decay chains including $\ell^{+}\ell^{-}$ and $Z$ boson, we choose the invariant mass larger than 12 GeV and $|M_{\ell\ell}-m_Z| > 10$ GeV.
\item
 Cut-2: At least two jets in the events are required to be successfully reconstructed, i.e., $N(j)\geq 2$. Among those reconstructed jets, there are at least one pair of jets which could come
from a $W$ boson either on-shell or off-shell. Thus the invariant mass of the $W$ boson is required  to be $M_{jj}<90$ GeV.
\item
 Cut-3: The invariant mass of $M_{\ell_{2}jj}$ is required to be smaller than 120 GeV.
\item
 Cut-4: Since the first lepton, $\ell_{1}$, is assumed to originate from the leptonically decaying top quark, the invariant mass of the $b$-jet and the leading lepton should be $M_{b\ell_{1}}< 140$ GeV.
\item
 Cut-5: The scalar sum of transverse momenta, $H_{T}$, is to be smaller than 250 GeV.
\end{itemize}

\begin{table}[htb]
\centering %
\caption{The cut flow of the cross sections (in fb) for the signal and SM backgrounds for the SS2L channel. The coupling parameters are taken as
$\kappa_{tuH}=0.1$ or $\kappa_{tcH}=0.1$ while fixing the other to zero. \label{cutflow-1}}
\vspace{0.2cm}
\begin{tabular}{p{2.0cm}<{\centering} p{1.5cm}<{\centering} p{1.5cm}<{\centering} p{1.8cm}<{\centering} p{0.5cm}<{\centering} p{1.5cm}<{\centering} p{1.5cm}<{\centering} p{1.5cm}<{\centering} p{1.0cm}<{\centering}}
\toprule[1.5pt]
\multirow{2}{*}{Cuts}& \multicolumn{3}{c }{Signal}&&\multicolumn{4}{c}{Backgrounds} \\ \cline{2-4}  \cline{6-9}
& $ug$ & $cg$ & $t\bar{t}\to tHq$ &&$t\bar{t}V$ & $WWjj$ & $WZjj$  & $t\bar{t}$ \\   \cline{1-9} \midrule[1pt]
Basic cuts & 3.12&0.34&3.77&&6.73&6.42&20.9&61004 \\ 
Cut 1 & 0.48&0.056&0.69&&0.85&0.21&0.25&6.52\\ 
Cut 2 &0.225&0.027&0.34&&0.27&0.04&0.046&2.54\\ 
Cut 3 & 0.18&0.022&0.28&&0.092&0.016&0.011&1.7\\ 
Cut 4 & 0.15&0.019&0.24&&0.058&0.009&0.0063&1.36\\ 
Cut 5 &0.14&0.017&0.21&&0.048&0.007&0.005&1.16\\ 
\bottomrule[1.5pt]
\end{tabular}
\end{table}

The effects of the  cuts on the signal and background processes are illustrated in Tab.~\ref{cutflow-1} for the SS2L channel, where the anomalous coupling parameters are taken as
$\kappa_{tuH}=0.1$ or $\kappa_{tcH}=0.1$ while fixing the other to zero.
 From Tab.~\ref{cutflow-1} we can see that, after all these cuts, the $t\bar{t}$ backgrounds
 for the SS2L channel, with fake leptons from heavy-flavor jets or charge mis-identifications
can be significant.

Obviously, the non-prompt backgrounds may also be significant, where non-prompt leptons are from heavy-flavor decays, mis-identified hadrons, muons from light-meson decays or electrons from un-identified conversions of photons into jets. Recently, the CMS collaboration searched for  SS2L signatures~\cite{epjc77-578} and found that the overall non-prompt backgrounds are about 1.5 times the $t\bar{t}W$ background after all cuts.  These non-prompt backgrounds are  not properly modeled in our MC simulations. Therefore, for simplicity, we add a non-prompt background that is
1.5 times $t\bar{t}W$~\cite{epjc77-578} after selection cuts to the overall background. Accounting for the theoretical and experimental systematic uncertainties on the background predictions would certainly improve the reliability of the results, yet they can only be neglected in our simulation.

\subsection{Analysis of the 3L channel}
Next, we consider the final states including 3L via the following processes:
\beq\label{signal}
pp&\to& t(\to W^{+}b\to \ell^{+}\nu b)h(\to WW^{\ast}\to \ell^{+}\nu \ell^{-}\bar{\nu}),\\
pp&\to& t(\to W^{+}b\to \ell^{+}\nu b)\bar{t}(\to Hq\to WW^{\ast}q\to \ell^{+}\nu \ell^{-}\bar{\nu}q),
\eeq
where $\ell =e, \mu$.

The dominant SM backgrounds are $t\bar{t}V$ ($V=W, Z$), $t\bar{t}H$, $WZ+$ jets and $t\bar{t}$. The multi-jet backgrounds~(where jets can fake electrons) are not included since they are negligible in multi-lepton analyses~\cite{14067830}.

The pre-selection cuts are taken as follows:
there must exist exactly three isolated leptons ($\ell=e, \mu$) and exactly one $b$-tagged jet with $p_{T}(\ell_{1}) > 20 \rm ~GeV$, $p_{T}(\ell_{2,3}) > 10 \rm ~GeV$, $p_{T}(j, b) > 20 \rm ~GeV$, $\slashed E_T> 100 \rm ~GeV$ and $|\eta_{\ell, j, b}|<2.5$.
 These cuts can strongly reduce the $t\bar{t}$ background and di-boson components.
\begin{figure}[htb]
\begin{center}
\vspace{0.5cm}
\centerline{\hspace{2.0cm}\epsfxsize=11cm\epsffile{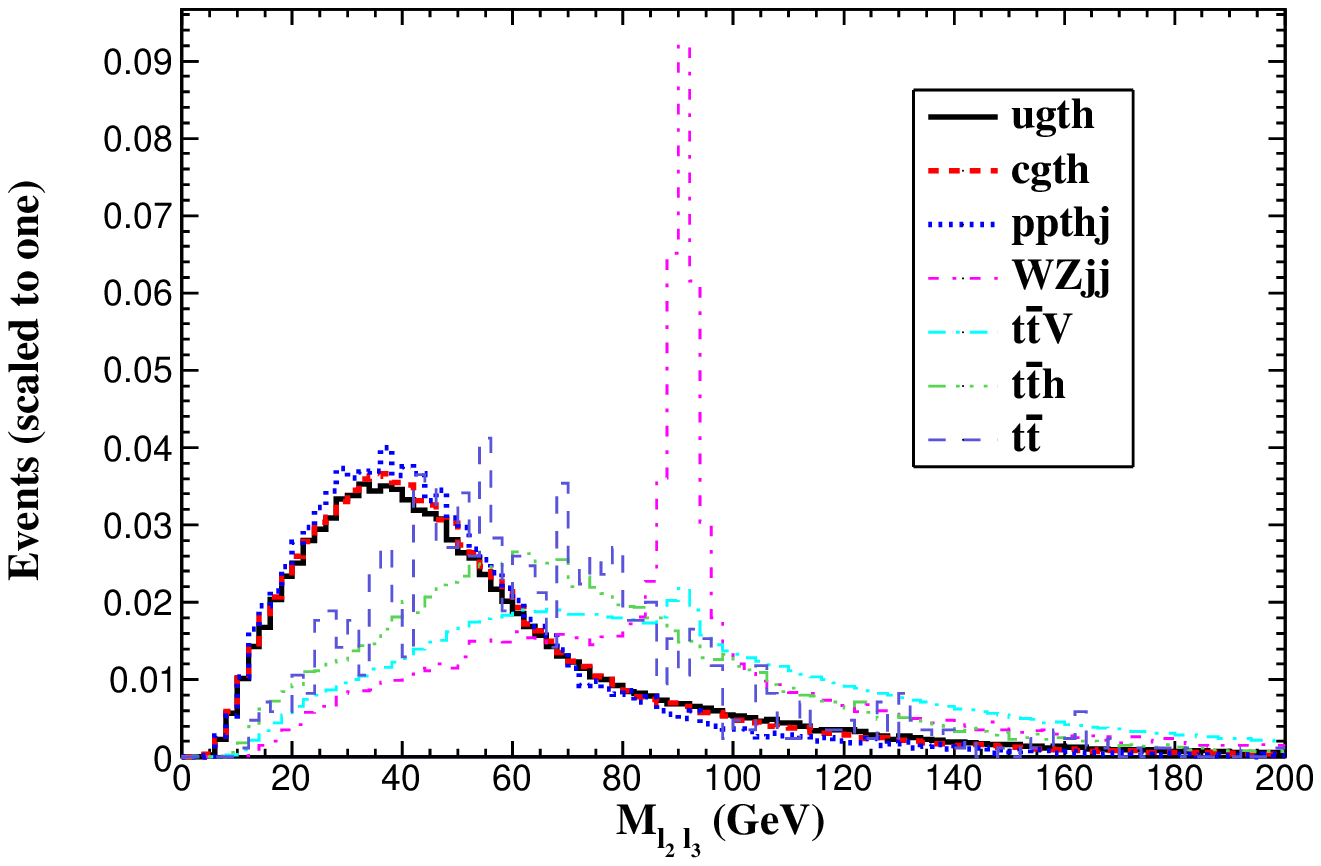}\hspace{-3.0cm}\epsfxsize=11cm\epsffile{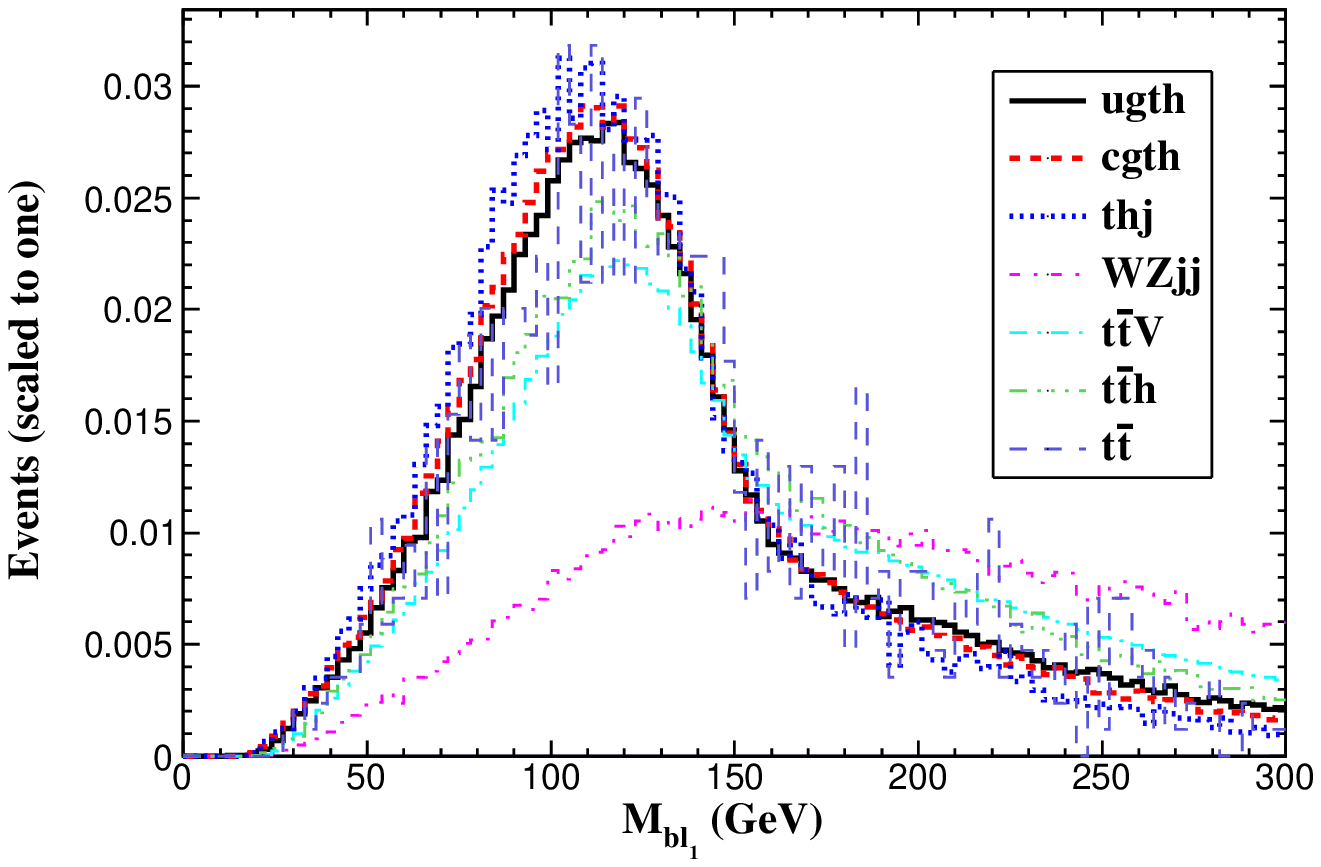}}
\caption{Normalized invariant mass distributions of $M_{\ell_{2}\ell_{3}}$ (left) and $M_{b\ell_{1}}$ (right).}
\label{mh3l}
\end{center}
\end{figure}

In Fig.~\ref{mh3l}, we show the invariant mass distribution of $M_{\ell_{2}\ell_{3}}$ and $M_{b\ell_{1}}$ from the signal and backgrounds at the 14 TeV LHC. To remove contamination from hadron decay chains including $\ell^{+}\ell^{-}$ pairs and  resonant $Z$ bosons, we choose the invariant mass $M_{\ell_{2}\ell_{3}}$ cuts
 \begin{itemize}
\item $12 {\rm GeV} < M(\ell_{2}\ell_{3})< 55 \rm ~GeV$.
\end{itemize}
Similarly, the invariant mass of the $b$-jet and the leading lepton, $M_{b\ell_{1}}$, should be smaller than 140 GeV. The effects of the cuts on the signal and backgrounds processes are illustrated in Table 2 for the 3L channel. One can see that significant backgrounds also come from the top pair production process with fake leptons or charge mis-identifications.

\begin{table}[htb]
\begin{center}
\caption{The cut flow of the cross sections (in fb) for the signal and background processes for the 3L channel. \label{cutflow-2}}
\vspace{0.2cm}
\begin{tabular}{p{2.0cm}<{\centering} p{1.5cm}<{\centering} p{1.5cm}<{\centering} p{2.0cm}<{\centering} p{1.5cm}<{\centering} p{1.5cm}<{\centering} p{1.5cm}<{\centering} p{2.0cm}<{\centering}}
\toprule[1.5pt]
\multirow{2}{*}{Cuts}& \multicolumn{3}{c}{Signals}&\multicolumn{4}{c}{Backgrounds} \\ \cline{2-8}
& $ug$ & $cg$ & $t\bar{t}\to tHq$ & $t\bar{t}$& $t\bar{t}V$ & $WZjj$  & $t\bar{t}h$  \\ \cline{1-8}
\midrule[1pt]
Basic cuts & 1.39&0.17&2.05&21843&1.85&46.2&0.025\\ 
After cuts & 0.14&0.018&0.106&0.23&0.024&0.021&$1.7\times 10^{-5}$\\ 
\bottomrule[2pt]
\end{tabular} 
\end{center}
\end{table}

\begin{figure}[htb]
\begin{center}
\vspace{-0.5cm}
\centerline{\epsfxsize=9cm \epsffile{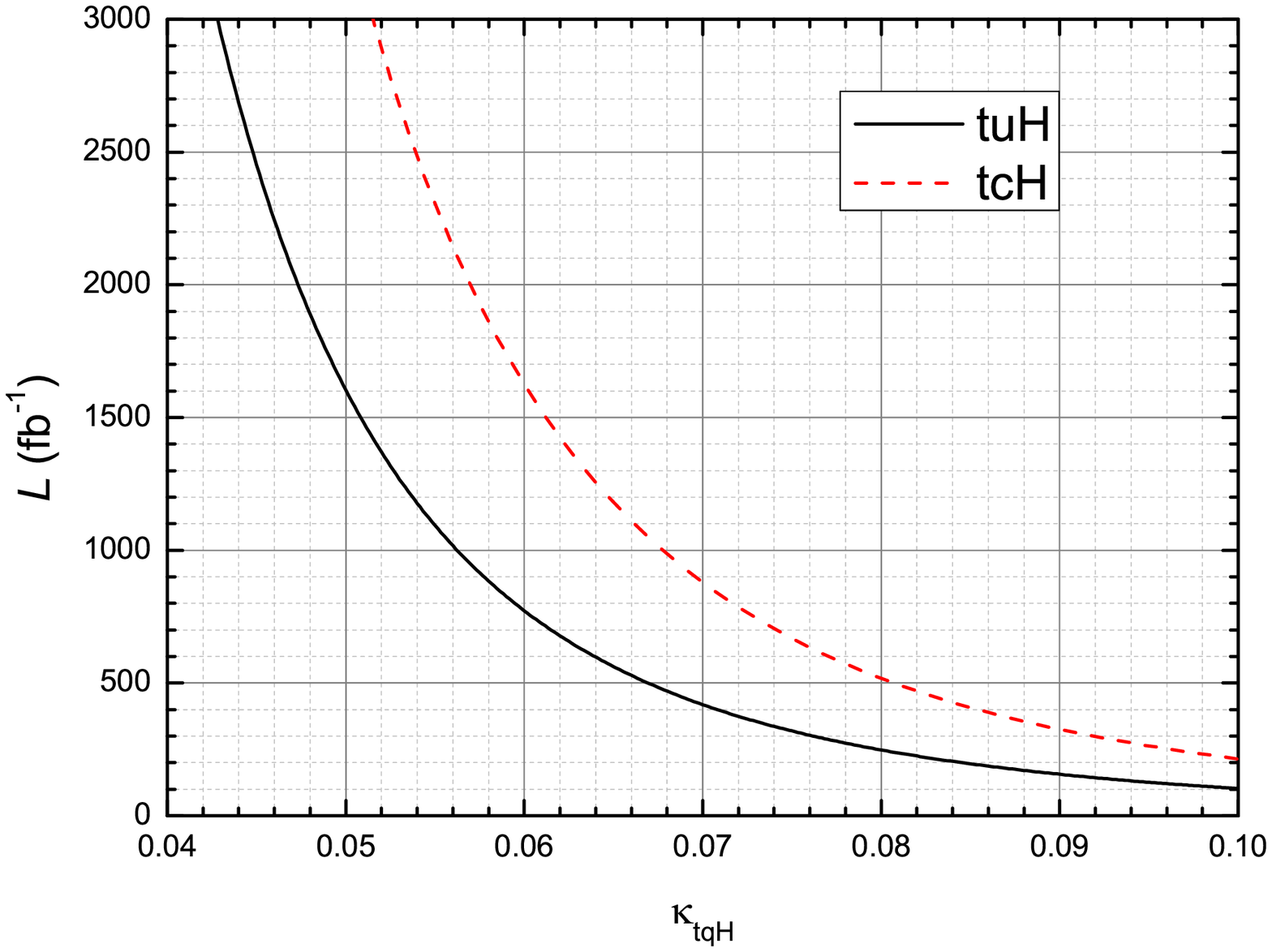}\hspace{-1.0cm}\epsfxsize=9cm \epsffile{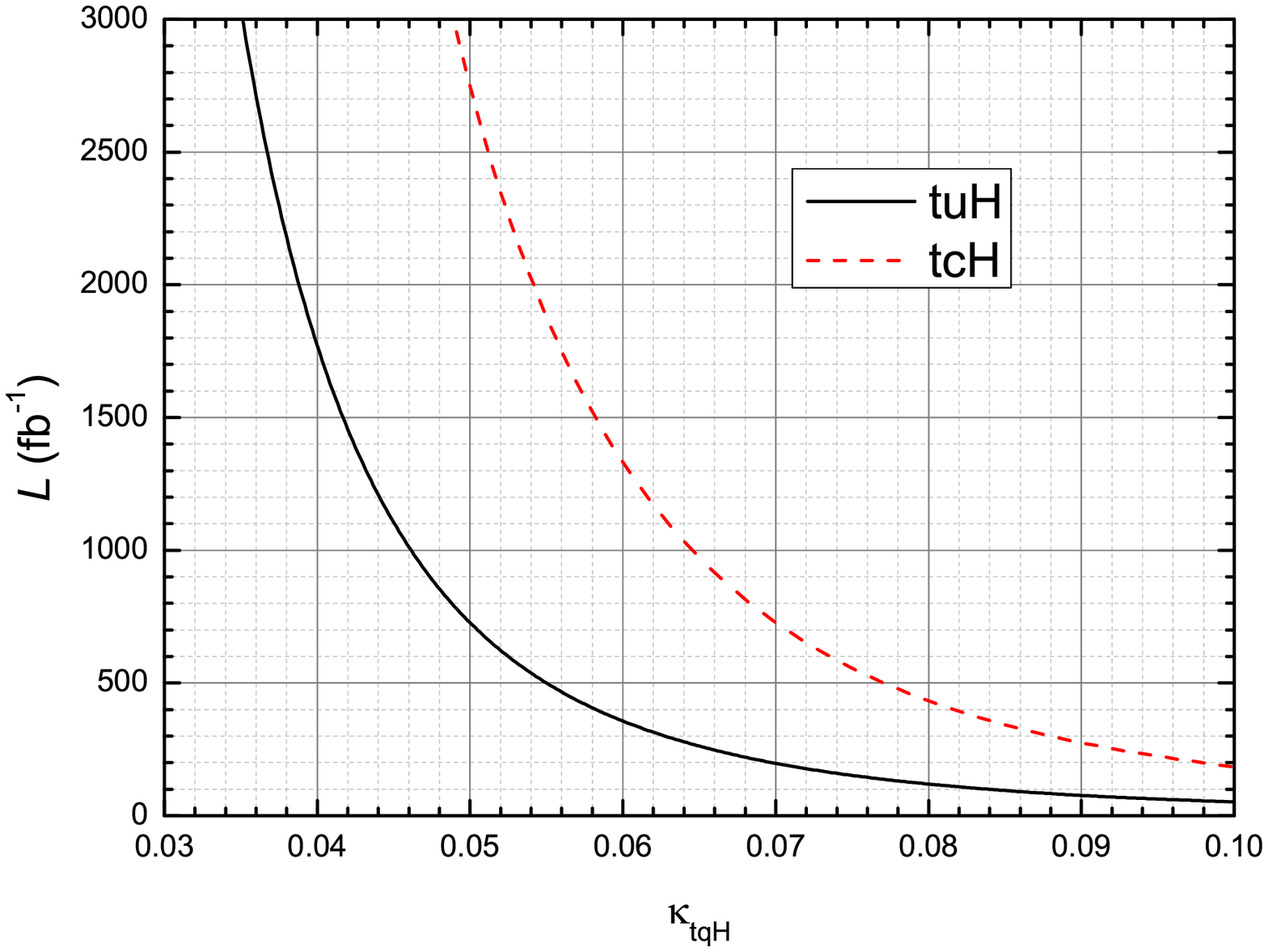}}
\caption{The $3\sigma$ contour plots for the signal in the ${\cal L}_{\rm int}-\kappa_{tqH}$ plane for the SS2L (left) and 3L (right) channels at the 14 TeV LHC. }
\label{ss3}
\end{center}
\end{figure}
\begin{figure}[htb]
\begin{center}
\vspace{-0.5cm}
\centerline{\epsfxsize=9cm \epsffile{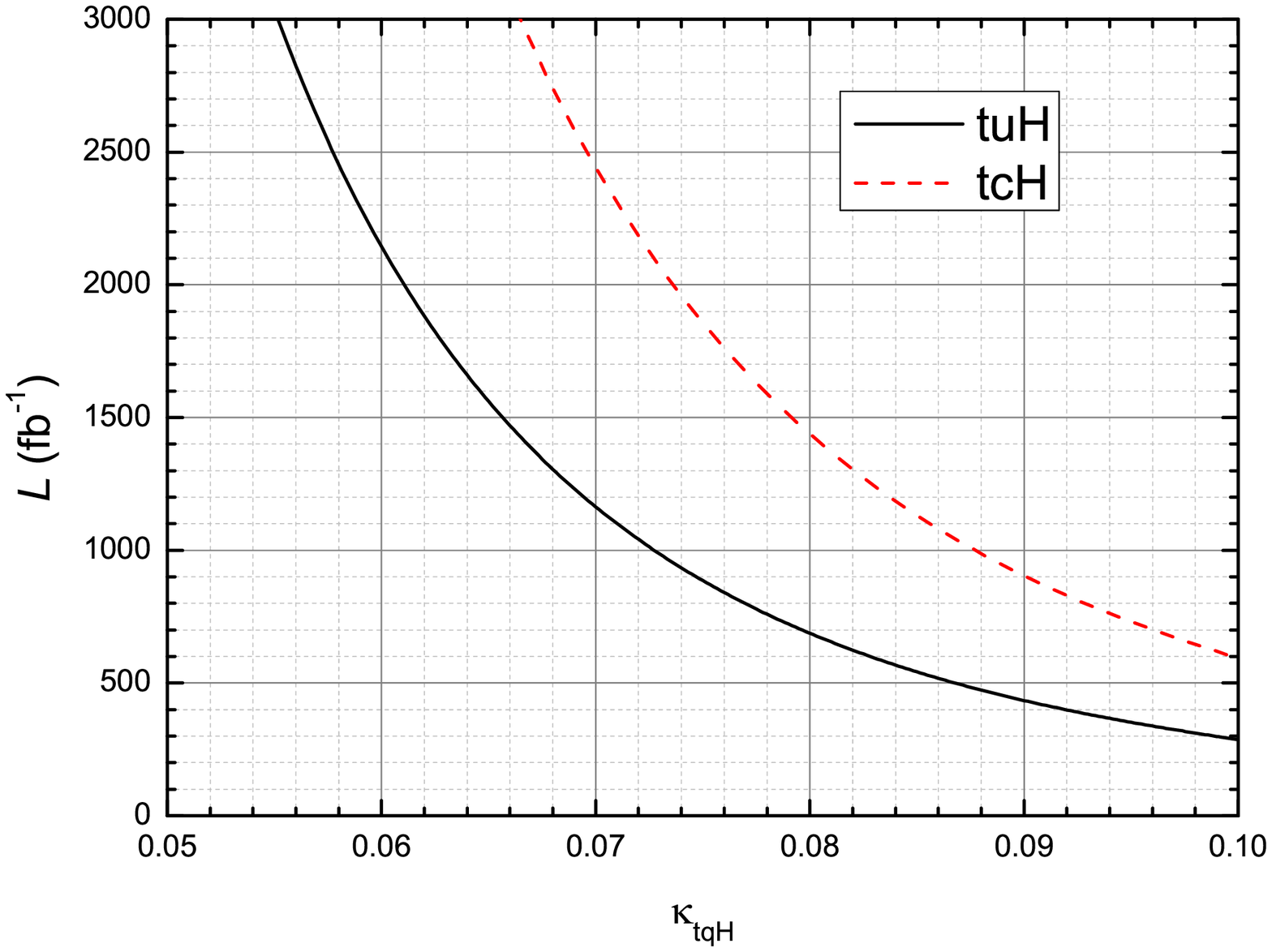}\hspace{-1.0cm}\epsfxsize=9cm \epsffile{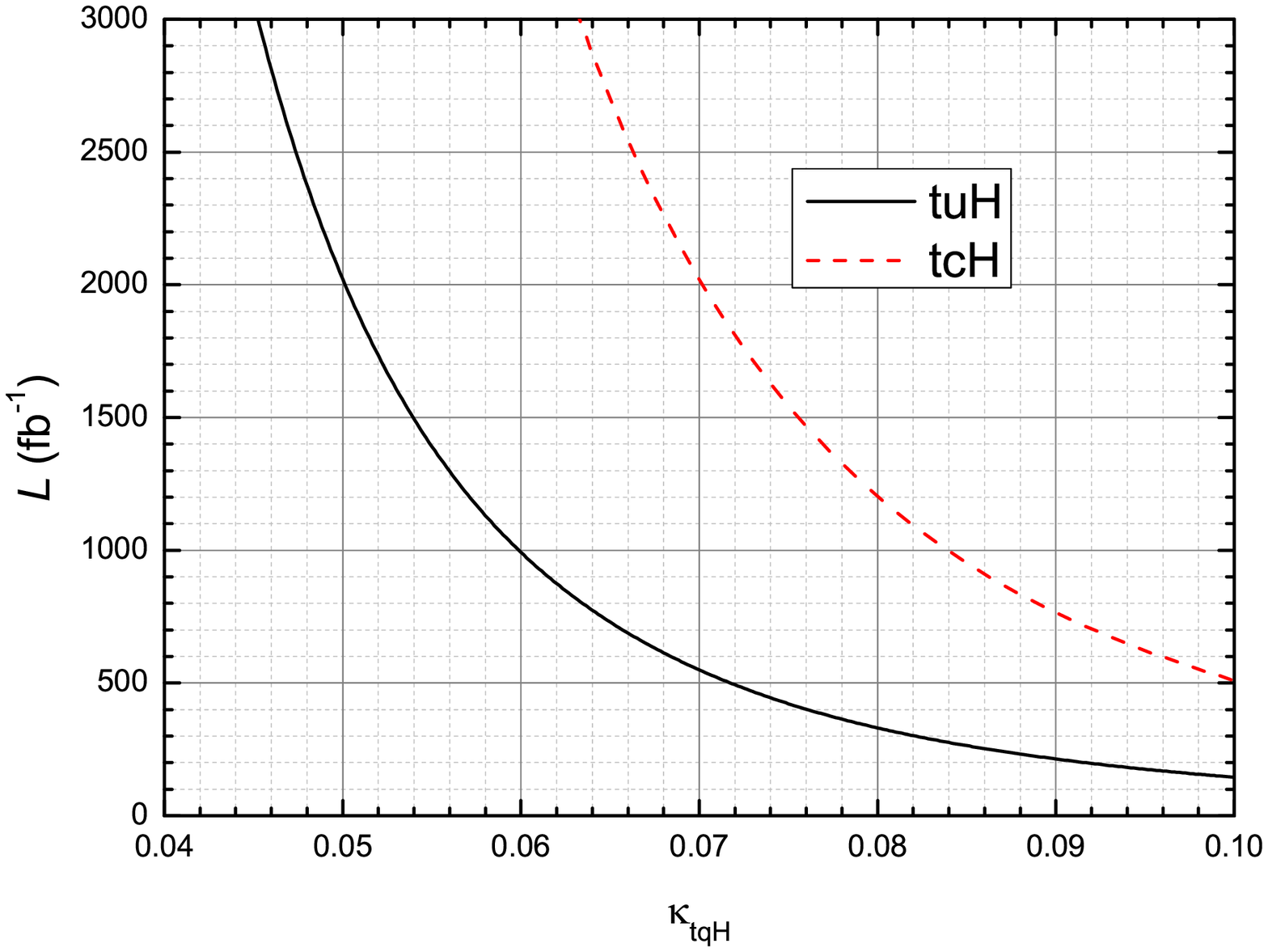}}
\caption{The $5\sigma$ contour plots for the signal in the ${\cal L}_{\rm int}-\kappa_{tqH}$ plane  for the SS2L (left) and 3L (right) channels at the 14 TeV LHC. }
\label{ss5}
\end{center}
\end{figure}
Using the Poisson formula
$SS=\sqrt{2{\cal L}_{\rm int}[(S+B)\ln(1+S/B)-S]}$~\cite{ss} we estimate the Signal Significance ($SS$) with fixed coupling parameters $\kappa_{tqH}$ and a given integrated luminosity ${\cal L}_{\rm int}$.
In Figs.~\ref{ss3} and \ref{ss5}, we plot the contours of $SS=3$ and $SS=5$, respectively, for two channels in the plane of ${\cal L}_{\rm int}-\kappa_{tqH}$. It is clear that, for an integrated luminosity of 3000 fb$^{-1}$, the FCNC couplings $\kappa_{tuH}~(\kappa_{tcH})$ can be probed to 0.045~(0.052) and 0.035~(0.049) at $3\sigma$ statistical sensitivity for the SS2L and 3L channels, respectively.
After neglecting the masses of light quarks, the branching ratio of $t \to qH$ is approximately given by \cite{jhep-1407-046,nlo}
\begin{equation}
Br(t \to qH) = \frac{\kappa^{2}_{tqH}}{\sqrt{2} m^2_t G_F}\frac{(1-x^2_h)^2}{(1-x^2_W)^2 (1+2x^2_W)}\lambda_{QCD} \simeq 0.58\kappa_{tqH}^{2},
\label{br}
\end{equation}
in terms of the Fermi constant $G_F$ and with $x_i=m_i/m_t~(i=W,\ h)$.
In our numerical calculation,
the relevant SM input parameters are taken as~\cite{pdg}:
\begin{align}
m_H&=125{\rm ~GeV}, \quad m_t=173.1{\rm ~GeV}, \quad m_W=80.379{\rm ~GeV},\\ \nonumber
m_Z&=91.1876{\rm ~GeV}, \quad \alpha_{s}(m_Z)=0.1185, \quad G_F=1.166370\times 10^{-5}\ {\rm GeV^{-2}}.
\end{align}
Using eq.~(\ref{br}), the limits can be translated in terms of constraints on the branching fractions of rare top decays. The $3\sigma$ CL upper limits on $Br(t\to qH)$ are about $Br(t\to uH)=1.17\times 10^{-3}$ and $Br(t\to cH)=1.56\times 10^{-3}$ for the SS2L channel, and $Br(t\to uH)=7.1\times 10^{-4}$ and $Br(t\to cH)=1.39\times 10^{-3}$ for the 3L channel. The projected limits from different
channels are summarized in Tab.~\ref{SBaftercuts_1}. We can see from the table that our results are comparable with the sensitivity limits at the HL-LHC as $Br(t\to uH)<0.036\%$ via the $H\to \gamma\gamma$ channel~\cite{t6}, $Br(t\to uH)<0.05\%$ via the multi-lepton channel and $Br(t\to uH)<0.02\%$ via the di-photon channel ~\cite{13112028}.

\begin{table}[htbp]
\renewcommand\arraystretch{0.9}
 \caption{\label{SBaftercuts_1}
The projected limits on $Br(t\to qH)$ from different  channels. The last two lines of the table are the results of this work.}
\vspace{-0.5cm}
\begin{center}
\scalebox{0.9}{\begin{tabular}{p{5.0cm}<{\centering}  p{8.0cm}<{\centering}  p{5.0cm}<{\centering} }
\toprule[1.5pt]
Channels      &Data Set &Limits  \\ \midrule[1pt]
$ tH\to \ell\nu b\tau^+\tau^-$ & LHC, 100 fb$^{-1}$ @ 13 TeV, $95\%$ CL  &$ Br$ $(t\to uH)< 0.15$ $\%$ \cite{jhep-1407-046}  \\
$ tH\to \ell\nu b\ell^+\ell^-X$ &LHC, 100 fb$^{-1}$ @ 13 TeV, $95\%$ CL  &$ Br$ $(t\to uH)< 0.22$ $\%$ \cite{jhep-1407-046} \\
$ t\bar{t}\to Wb + Hc\to jj b +\tau\tau c$ &LHC, 100 fb$^{-1}$ @ 13 TeV, $95\%$ CL  &$ Br$ $(t\to cH)< 0.25$ $\%$ \cite{150908149}\\
$ tH\to jjb b\bar{b}$ &LHC, 100 fb$^{-1}$ @ 13 TeV, $95\%$ CL  &$ Br$ $(t\to uH)< 0.36$ $\%$ \cite{jhep-1407-046}\\
$ Wt\to WHq \to \ell\nu b\gamma\gamma q$ & LHC, 3000 $ fb^{-1}$ @ 14 TeV, $3\sigma$ &   $ Br$ $(t\to qH)< 0.24$ $\%$ \cite{t5}         \\
$ tH\to \ell\nu b\gamma\gamma q$ & LHC, 3000 $ fb^{-1}$ @ 14 TeV, $3\sigma$ &   $ Br$ $(t\to uH)< 0.036$ $\%$ \cite{t6}         \\
$ t\bar{t}\to WbqH\to \ell\nu b\gamma\gamma q$  &  LHC, 3000 $ fb^{-1}$ @ 14 TeV, $3\sigma$     &   $ Br$ $(t\to uH)< 0.23$ $\%$ \cite{t7}       \\
$ e^{-}p\to \nu_{e}\bar{t}\to \nu_{e}H(\to b\bar{b}) \bar{q}$ & LHeC, 200 $ fb^{-1}$ @ 150 GeV $\oplus$ 7 TeV, $95\%$ CL &   $ Br$ $(t\to qH)< 0.013$ $\%$ \cite{t8}        \\
$ t\bar{t}\to tqH\to \ell\nu bb\bar{b} q$ &   ILC, 3000 fb$^{-1}$ @ 500 GeV, $95\%$ CL       &   $ Br$ $(t\to qH)< 0.112$ $\%$ \cite{t9}        \\
$ t\bar{t}\to tqH\to \ell\nu bb\bar{b} q$ &   ILC~(unpolarized), 500 fb$^{-1}$ @ 500 GeV, $3\sigma$      &   $ Br$ $(t\to qH)< 0.119$ $\%$ \cite{t10}        \\
$ t\bar{t}\to tqH\to \ell\nu bb\bar{b} q$ &   ILC~(polarized), 500 fb$^{-1}$ @ 500 GeV, $3\sigma$       &   $ Br$ $(t\to qH)< 0.088$ $\%$ \cite{t10}        \\
$ t\bar{t}\to Wb + Hq\to \ell \nu b +\gamma\gamma q$ &LHC, 3000 fb$^{-1}$ @ 14 TeV, $95\%$ CL  &$ Br$ $(t\to qH)< 0.02$ $\%$ \cite{13112028}\\
$ t\bar{t}\to Wb + hq\to \ell \nu b +\ell\ell qX$ &LHC, 3000 fb$^{-1}$ @ 14 TeV, $95\%$ CL  &$ Br$ $(t\to qH)< 0.05$ $\%$ \cite{13112028}\\
\multirow{2}{*}{This work for the SS2L channel}&\multirow{2}{*}{LHC, 3000 fb$^{-1}$ @ 14 TeV, $3\sigma$ }&$ Br$ $(t\to uH)< 0.117$ $\%$, \\
&& $ Br$ $(t\to cH)< 0.156$ $\%$ \\ 
\multirow{2}{*}{This work for the 3L channel}&\multirow{2}{*}{LHC, 3000 fb$^{-1}$ @ 14 TeV, $3\sigma$ }&$ Br$ $(t\to uH)< 0.071$ $\%$, \\
&& $ Br$ $(t\to cH)< 0.139$ $\%$ \\
\bottomrule[1.5pt]
 \end{tabular}}
 \end{center}
\end{table}
\section{Conclusions}
The discovery of the 125 GeV Higgs boson opens the door to probe NP processes that involve Higgs boson associated production or decay. In this paper, we have investigated the signal of  $tH$ associated production via  FCNC $tqH$ couplings and $t\bar{t}$ production with $\bar{t}\to H\bar{q}$ at the 14 TeV LHC. We focused on the final states including SS2L and 3L signals from the decay modes $t\to b\ell^{+}\nu_{\ell}$, $H\to WW^{\ast}\to \ell^{+}\nu jj $ or $H\to WW^{\ast}\to \ell^{+}\nu \ell^{-}\bar{\nu}$. We have then shown that, at $3\sigma$ level, the branching ratios $Br(t\to uH)$ and $Br(t\to cH)$ are, respectively, about $Br(t\to uH) \leq 1.17\times 10^{-3}$ and $Br(t\to cH) \leq 1.56\times 10^{-3}$ for the SS2L channel, and $Br(t\to uH) \leq 7.1\times 10^{-4}$ and $Br(t\to cH) \leq 1.39\times 10^{-3}$ for the 3L channel at the future HL-LHC.

\begin{acknowledgments}
The work of Y-B Liu is supported by the Foundation of Henan Institute of Science and Technology (Grant no. 2016ZD01) and the China Scholarship Council (201708410324). The work of SM is supported in part by the NExT Institute and the STFC CG ST/L000296/1.
\end{acknowledgments}

\end{document}